# Uncovering the Skillsets Required in Computer Science Jobs Using Social Network Analysis

Mehrdad Maghsoudi

*Department of Industrial and Information Management, Faculty of Management and Accounting, Shahid Beheshti University, Tehran, Iran, Email:* M_Maghsoudi@Sbu.ac.ir

## Abstract

the rapid growth of technology and computer science, which has led to a surge in demand for skilled professionals in this field. The skill set required for computer science jobs has evolved rapidly, creating challenges for those already in the workforce who need to adapt their skills quickly to meet industry demands. To stay ahead of the curve, it is essential to understand the hottest skills needed in the field. The article introduces a new method for analyzing job advertisements using social network analysis to identify the most critical skills required by employers in the market.

In this research, to form the communication network of skills, first 5763 skills were collected from the LinkedIn social network, then the relationship between skills was collected and searched in 7777 computer science job advertisements, and finally, the balanced communication network of skills was formed. The study analyzes the formed communication network of skills in the computer science job market and identifies four distinct communities of skills: Generalists, Infrastructure and Security, Software Development, and Embedded Systems. The findings reveal that employers value both hard and soft skills, such as programming languages and teamwork. Communication skills were found to be the most important skill in the labor market. Additionally, certain skills were highlighted based on their centrality indices, including communication, English, SQL, Git, and business skills, among others. The study provides valuable insights into the current state of the computer science job market and can help guide individuals and organizations in making informed decisions about skills acquisition and hiring practices.

# 1. Introduction

The world today is rapidly becoming technology-oriented, and this shift is evident in almost every aspect of our lives. Whether we're communicating with friends and family or shopping for goods and services, computers and the internet have become integral to our daily routines(Alkureishi et al., 2021). This transformation has been driven by computer science, which has played a significant role in creating innovative solutions to complex problems, opening up possibilities that were once considered impossible(Javed et al., 2022).

As the growth of computer science continues to outpace traditional industries, there has been a surge in demand for skilled professionals in this field(Ainslie & Huffman, 2019). The skill set required for computer science jobs has evolved rapidly, with an increasing focus on advanced programming languages, analytics, and data science(Verma et al., 2022). Keeping pace with these changes can be challenging, particularly for those already in the workforce who may need to adapt their skills to meet the demands of this high-speed industry(Su et al., 2021).

The fast-paced nature of technological change can pose a challenge to the workforce, particularly those who need to adapt their skills to meet the demands of the industry quickly (Doherty & Stephens, 2023). Finding and promoting a career in computer science requires not only technical expertise but also a willingness to adapt, learn, and innovate (Pournaghshband & Medel, 2020). As the job market becomes increasingly competitive, it is essential to stay ahead of the curve by understanding the hottest skills needed in the field. Training course organizers must keep pace with the latest trends and technologies to ensure they offer the most relevant and up-to-date training programs (Snell & Morris, 2022).

One area that showcases the power of computer science is social network analysis. With this technology, we can analyze large-scale data sets to identify patterns, connections, and relationships between individuals or groups, providing insights into macro-level events (Kermani et al., 2022). By leveraging this technology, we can better understand the changing demands of the industry and help job seekers and training course organizers set more precise goals that match market needs.

To accomplish this goal, our article introduces a new method for analyzing job advertisements and identifying the most essential and critical skills required by employers in the market. By using social network analysis, we can gain insights into the skills required for success in computer science careers. This method offers a practical approach that can be used by both job seekers and training course organizers to stay ahead of the curve and give themselves a competitive edge in this ever-evolving field.

Continuing with this article, we will move on to the second chapter where we will present research literature and provide a summary of related works. The third chapter will focus on discussing the methodology used for conducting this study and detail the process of collecting the required data. The fourth section will comprehensively discuss the research results obtained. Lastly, the fifth chapter will be dedicated to presenting our conclusions based on the findings.

# 2. Literature review

## 2.1. Social Network Analysis

Social Network Analysis (SNA) is a technique that has gained popularity in recent times, especially with the rise of social media platforms. It refers to the study of social networks and the relationships between individuals or groups within them. The idea behind SNA is to analyze patterns of communication, collaboration, and information flow among people or organizations(Kermani et al., 2022).

The basic unit of analysis in SNA is the node, which represents an individual or group. Nodes are connected through ties, which represent various types of relationships such as friendships, collaborations, and communication channels. These ties can be visualized using network diagrams, which provide a graphical representation of the network structure (Maghsoudi, Jalilvand Khosravi, et al., 2023).

One of the main advantages of SNA is that it allows researchers to gain insights into the structure and dynamics of social networks(Kurt & Kurt, 2020). By analyzing the connections between nodes, researchers can identify key actors, communities, and subgroups within the network(Kermani et al., 2022). They can also study the flow of information or influence within the network and determine how it affects the behavior of individuals or groups(Xu et al., 2023).

SNA has a wide range of applications in various fields, including sociology, psychology, anthropology, political science, and business(Su et al., 2020). In sociology, SNA can be used to study social movements, organizational behavior, and social capital(Packer et al., 2020). In psychology, it can be applied to understand the spread of emotions and attitudes(Alqahtani & Alothaim, 2022). In business, SNA can help organizations identify influential customers or suppliers and improve their communication and collaboration strategies(Toth et al., 2021).

There are several methods for conducting SNA, including centrality measures, community detection, and network visualization. Centrality measures refer to the calculation of measures such as degree centrality, betweenness centrality, and closeness centrality, which help to identify the most important nodes in the network. Community detection involves identifying subgroups within the network based on the patterns of connections between nodes. Network visualization provides a way to visualize the network structure, making it easier to understand and interpret(Maghsoudi, Shokouhyar, et al., 2023).

Social Network Analysis is a powerful technique that provides insights into the structure and dynamics of social networks(Ryan & Dahinden, 2021). It has numerous applications in various fields and can help us understand the behavior of individuals and groups within these networks. By using SNA, researchers can gain valuable insights that can inform policy-making(Varone et al., 2017), organizational strategy(Guevara et al., 2020), and social interventions(Smit et al., 2020).

### 2.1.1. Social network analysis indicators

**Node**: A node is a fundamental unit of analysis in social network analysis (Gou & Wu, 2022). It represents an individual, group, or entity within a network. Nodes can be anything from a person, an organization, a geographic location, or even a concept. In social network analysis, nodes are used to represent the actors involved in a network and their interactions with one another (Chen et al., 2020).

**Edge**: An edge (also called a tie, link, or connection) is a line connecting two nodes in a network. It represents a relationship or interaction between the nodes it connects. Edges can be directed (when the relationship has a specific direction) or undirected (when the relationship is symmetrical). In social network analysis, edges are used to study the patterns of communication, collaboration, and information flow among nodes (Kermani et al., 2022).

**Degree**: The degree of a node refers to the number of edges connected to that node. It represents the number of direct connections that a node has within a network. In social network analysis, the degree is a commonly used measure of centrality and helps to identify the most important nodes in a network (Lotf et al., 2022).

**Density**: The density of a network refers to the proportion of possible edges that are present in the network. It provides a measure of how closely connected the nodes are in the network. High density indicates that many of the possible edges are present, while low density indicates that fewer edges are present. In social network analysis, density is used to assess the overall structure and connectivity of a network (Maghsoudi, Shokouhyar, et al., 2023).

**Centrality Measures**: Centrality measures are used to quantify the importance of a node in a network. The most used centrality measures include (HabibAgahi et al., 2022):

- Degree centrality: The degree of a node, i.e., the number of edges connected to that node.
- Betweenness centrality: The extent to which a node lies on the shortest path between other nodes in the network.
- Closeness centrality: The reciprocal of the sum of the shortest distances from a node to all other nodes in the network.
- Eigenvector centrality: The importance of a node is based on the centrality of the nodes it is connected to.

### 2.1.2. Community in Social Networks Analysis

The real-world community is typically a group of individuals with similar economic, social, or political interests who often reside nearby. Virtual communities, on the other hand, form when users connect through social media and engage with each other. To constitute a community, there must be at least two connections sharing a common interest and commitment to that interest (Zafarani et al., 2014). A community can be defined as a group of entities that are closer to one another than to other entities in the dataset (Alhayan et al., 2023). These groups emerge when people interact more frequently within a group than outside of it (Figure 1). The proximity of entities within a community can be measured by examining the similarity or distance between them. In essence, a social network community is akin to a network cluster (Maghsoudi, Shokouhyar, et al., 2023).

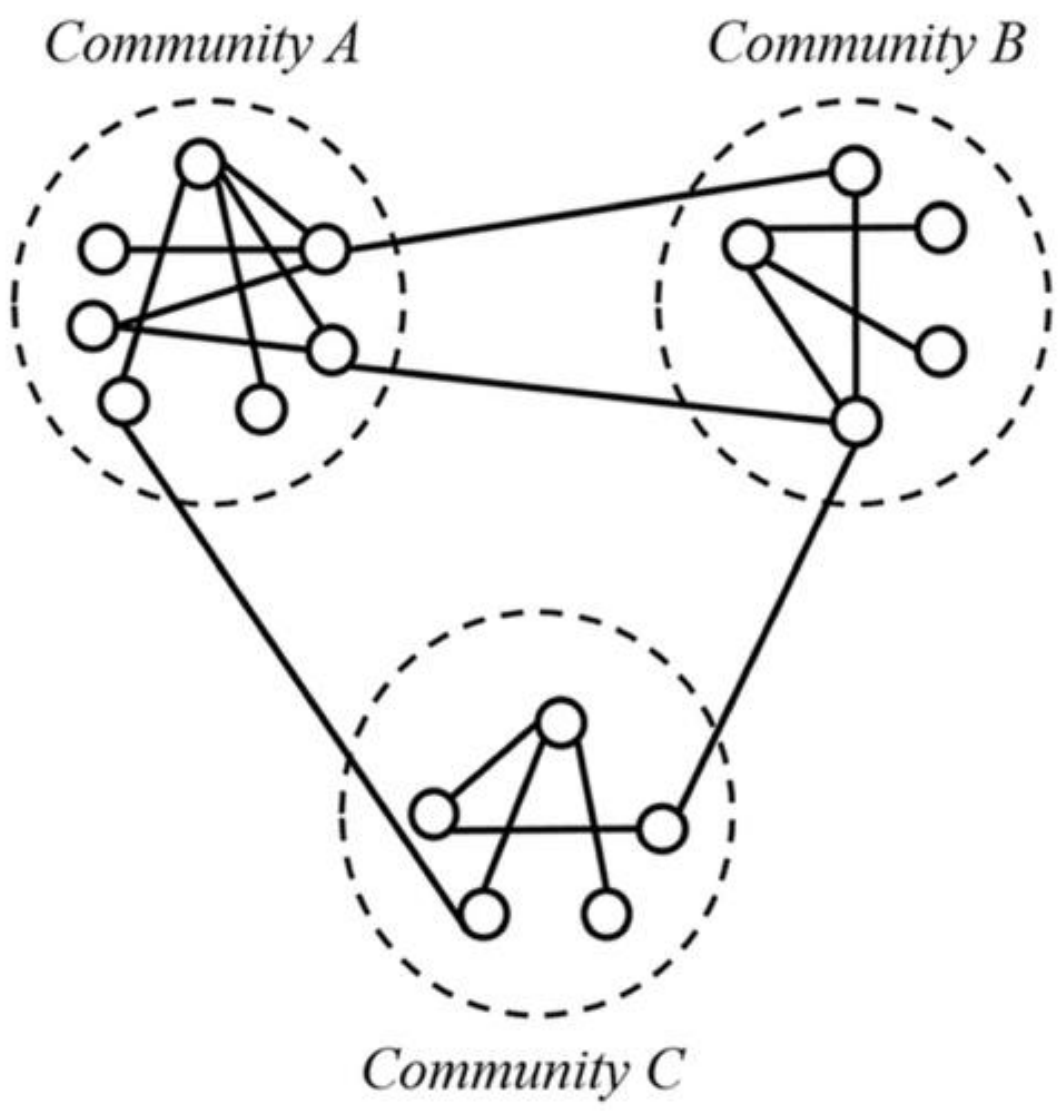

*Figure 1: Communities clustering (Ding et al., 2021)*

Community detection in network models is achieved through clustering, as illustrated in Figure 2. Communities are defined as groups of closely associated nodes with stronger connections to each other than to nodes in other communities (Tanhapour et al., 2022). Community detection is a valuable tool for network analysts to understand the interactions and cohesive sub-groups within a network. Modularity, as defined by (Newman, 2006), measures the performance of social network clustering. The algorithm for community detection in a weighted network with n nodes first considers each node as a community. It then finds a neighboring community for each node that maximizes the modularity index when the node is moved. If the move increases modularity, the node is moved to the new community; otherwise, it remains in its original community. This process is repeated for all nodes until no more changes are applied and a locally optimal point is reached. In the second phase of the algorithm, small communities are merged to create larger ones until the maximum modularity index is achieved. Figure 2 shows how the algorithm identifies four communities in the first phase and merges them into two in the second phase to achieve the maximum modularity index (Yap et al., 2019).

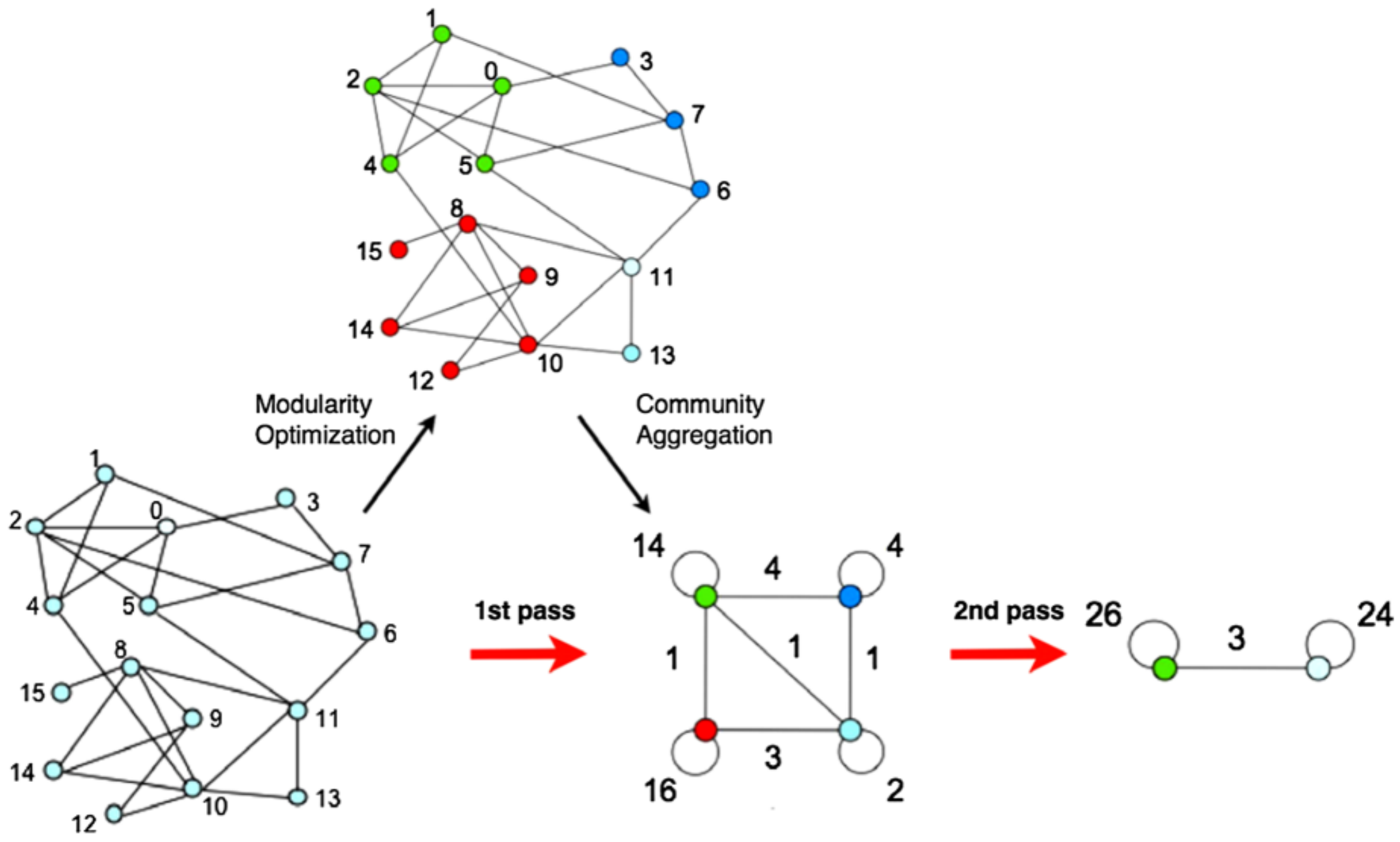

*Figure 2: Community detection based on increasing modularity(Yap et al., 2019)*

## 2.2. Related Works

In this section, we discuss the efforts made by researchers to analyze and predict the available capabilities in the job market and job advertisements.

(Mohammad Akhriza et al., 2017) developed solutions to help ICT students assess the gap between their skills and the rapidly evolving job skills required by the industry. They introduced measures and visualization tools using evolutionary-based data mining, collecting skillsets mastered by students from study reports and frequent skillsets required by the industry from job adverts to approximate student skill coverage. The proposed solutions were tested on data from an Indonesian higher education institution that implements a competency-based curriculum. Experimental results showed that the proposed approaches not only revealed and visualized the gap but also monitored changes in skill requirements, helping school administrators update the curriculum accordingly.

(Cicek et al., 2019) analyzed the future skill requirements in the maritime industry to address the changes necessary due to Industry 4.0 and emerging technologies. They examine the gaps between the training offered and the actual needs of the industry and comprehensively analyze the future skill requirements. The outcomes of the paper can be used by stakeholders in the maritime industry to take relevant actions regarding the impact of Industry 4.0.

(Gurcan & Cagiltay, 2019) propose a methodology for identifying the competencies required for big data software engineering by analyzing job advertisements using latent Dirichlet allocation (LDA). They aim to develop a taxonomy that maps these competencies, which could be used to evaluate and improve IT professionals' knowledge, identify professional roles in personnel recruitment, and meet industry skill requirements through education programs. The proposed model could also be extended to other online sources to identify emerging trends and generate contextual tags.

(Verma et al., 2019) conducted a content analysis of online job postings for business analyst, business intelligence analyst, data analyst, and data scientist positions. They ranked relevant skills for each position and compared the skill sets of data analysts vs. data scientists and business analysts vs. business intelligence analysts. The analysis identified decision-making, organization, communication, and structured data management as important across all positions. Technical skills, such as statistics and programming, were found to be most in demand for data analysts. These findings are useful for defining required skills for job categories and designing course curricula in the business and data analytics domain.

(Srinivasan & Thangaraj, 2021) conducted a study to identify the soft skills required by management graduates in India with a specialization in finance to become employable for finance job roles. They collected data from 117 finance professionals with a minimum work experience of 5 years and used factor analysis to identify 15 essential soft skills, such as empathy, ethical behavior, problem-solving, and effective communication, among others. The findings will help prospective candidates prepare themselves for a career in finance by developing the necessary soft skills apart from technical knowledge and hard skills.

(Aljohani et al., 2022) conducted a bibliometric analysis of cross-disciplinary records in Scopus to analyze the discussions on "curriculum alignment" in the context of "learned skills" and "acquired skills". They collected 10,214 data points from 2010 to 2021 to identify issues, names, and themes that have contributed to the field over the past decade. The research highlights the value and application of bibliometric analyses and emphasizes the need to ensure the compatibility of various scientific information archives to facilitate practical debate between academia and policymakers.

(Alibasic et al., 2022) The authors developed a data-driven approach to identify trending jobs and changes in the job market, focusing on the oil and gas industry. They used a range of data analytics tools such as LSI, LDA, Factor Analysis, and NMF to identify disparities between skills covered by the education system and those required in the job market. The study showed that while low-skilled jobs are at risk of being replaced by automation, some high-skilled jobs may also be affected. The findings emphasize the need for decision-makers to prepare the workforce for highly demanding jobs and the new skills required for collaboration between humans and machines.

(van Heerden et al., 2023) investigated and compared the soft skills possessed by industry professionals to those required by the industry to reduce sector-wide turnover. A questionnaire was administered to 741 respondents, and descriptive statistics and principal component factor analysis were used to analyze and classify the identified soft skills. Three clusters of possessed soft skills and two clusters of required soft skills were identified. The study concludes that certain soft skills are more controllable and easier to train than others, which are trait-based and more difficult to train and possess. Overall, the findings can help construction sector professionals plan for the appropriate skills they require and mitigate turnover. The summary of the related works and their respective implementation methods are presented in Table 1.

*Table 1: The summary of the related works*

| Year | Topic | Method |
|---|---|---|
| 2017 | Assessing skill gaps between students and industry requirements in ICT | Evolutionary-based data mining, visualization tools |

| Year | Topic | Method |
|---|---|---|
| 2019 | Analyzing future skill requirements in the maritime industry | Gap analysis, comprehensive future skill analysis |
| 2019 | Identifying competencies required for big data software engineering | Latent Dirichlet allocation (LDA), taxonomy mapping |
| 2019 | Ranking relevant skills for business and data analytics job positions | Content analysis of online job postings |
| 2021 | Identifying soft skills required by finance professionals in India | Factor analysis |
| 2022 | Bibliometric analysis of curriculum alignment discussions | Bibliometric analysis |
| 2022 | Identifying trending jobs and changes in the oil and gas industry | Data analytics tools, such as LSI, LDA, Factor Analysis, and NMF |
| 2023 | Comparing soft skills possessed by industry professionals to those required by the industry in the construction sector | Questionnaire, descriptive statistics, principal component factor analysis |

This article stands out for its unique approach to analyzing the skillsets required in computer science jobs using social network analysis. While previous studies have focused on identifying gaps in skills or predicting future skill requirements, this research utilizes a communication network analysis to uncover the most important skill requirements of the computer science job market. By applying the index centrality to analyze the communities within the communication network, this study provides valuable insights into which skills are most essential for success in the computer science field. This information can be used by educators to design more effective curricula and by job seekers to understand which skills they need to develop to succeed in their careers.

## 3. Research Methodology

Figure 3 mentions the methodology and the general direction of this research.

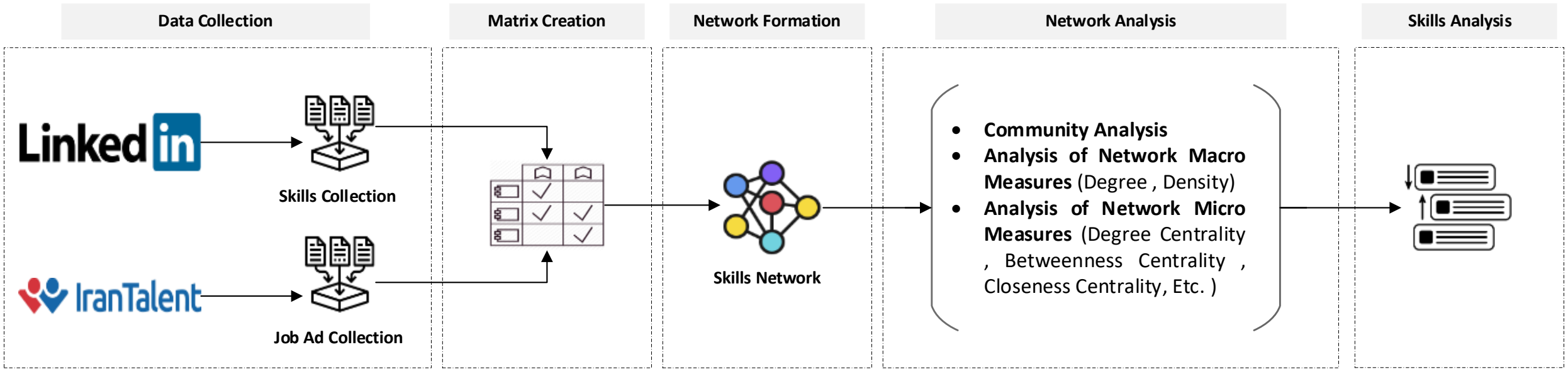

Figure 3: Research methodology

**Collecting data**:

The data collection stage consists of two parts:

A) **Collecting the list of skills**:

LinkedIn is the largest professional network in the world (Mazurek et al., 2022). This social network has collected a large collection of skills and expertise, and for this reason, it is a suitable source for collecting the list of skills. In this research, we collect the list of skills from LinkedIn using the Python programming language and Request package.

B) **Collecting job ads**:

To collect job advertisements in this research, we use the "Iran Talent" website. Iran Talent is the largest and oldest job site in Iran and one of the largest job sites in the Middle East (Salamzadeh & Ramadani, 2021). Data collection of job ads is also done using Python programming language.

**Data preprocessing:**

In the data preprocessing stage, the goal is to construct the Ad-skill matrix. This matrix shows which of the skills are present in each advertisement, and the output of this matrix is a table similar to Table 2. This table is used to build the skills network.

*Table 2: Ad-skill matrix*

| Job Ad | Skill 1 | Skill 2 | Skill 3 | Skill 4 | … | Skill n |
|---|---|---|---|---|---|---|
| Job Ad 1 | 0 | 1 | 1 | 0 | … | 1 |
| Job Ad 2 | 2 | 1 | 0 | 1 | … | 0 |
| … | … | … | … | … | … | … |
| Job Ad n | 0 | 0 | 1 | 2 | … | 0 |

**Network Formation**:

Based on the matrix created in the previous step, the work of building the communication network of the skills is done. Based on this, the basis of the relationship between the two skills is their joint presence in a job advertisement. Accordingly, the more two skills have a common presence in job ads, the more weight the link between them will have. Figure 4 shows the interface of establishing communication between skills in the network.

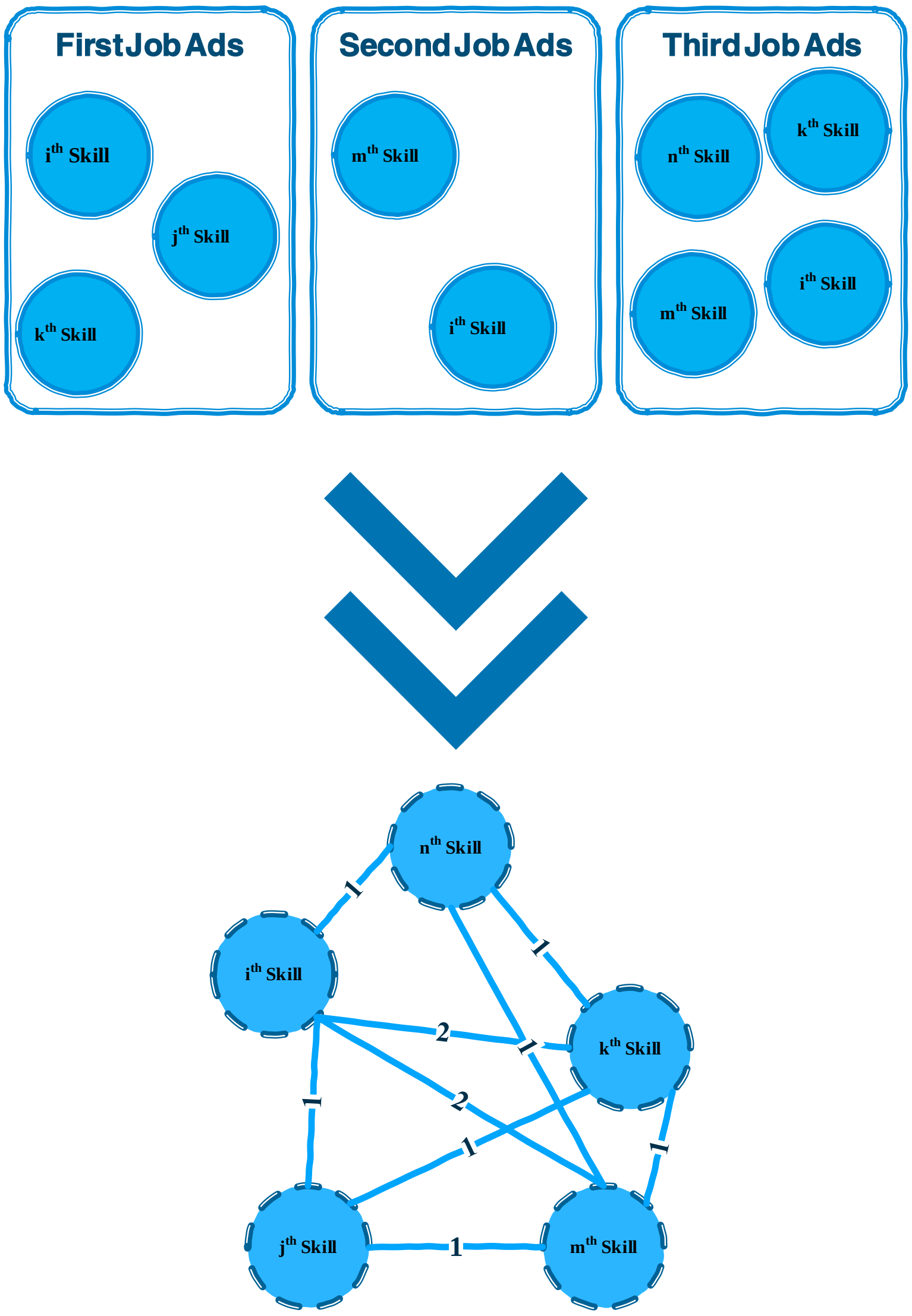

*Figure 4:Skill network formation*

**Network Analysis**:

In our network analysis, we use different indicators of social network analysis. In addition to expressing general indicators such as the degree of skills and network density, we analyze the importance of each of the skills using centrality indicators and choose the most basic skills of the job advertisements. Also, by analyzing the existing communities of the skills network, we will categorize the types of skills in the network.

**Skills analysis**:

In the last step, using the set of analyzes available in the "Network Analysis" step, we summarize the importance and community of the skills.

# 4. Results

## 4. 1 Data gathering

After extracting skills from the LinkedIn social network, a total of 7345 skills were identified. However, after eliminating duplicate entries, the number was reduced to 5763. Figure 5 shows a representation of the set of crawled skills. Additionally, 7350 job listings about computer science were discovered during the Crawley search.

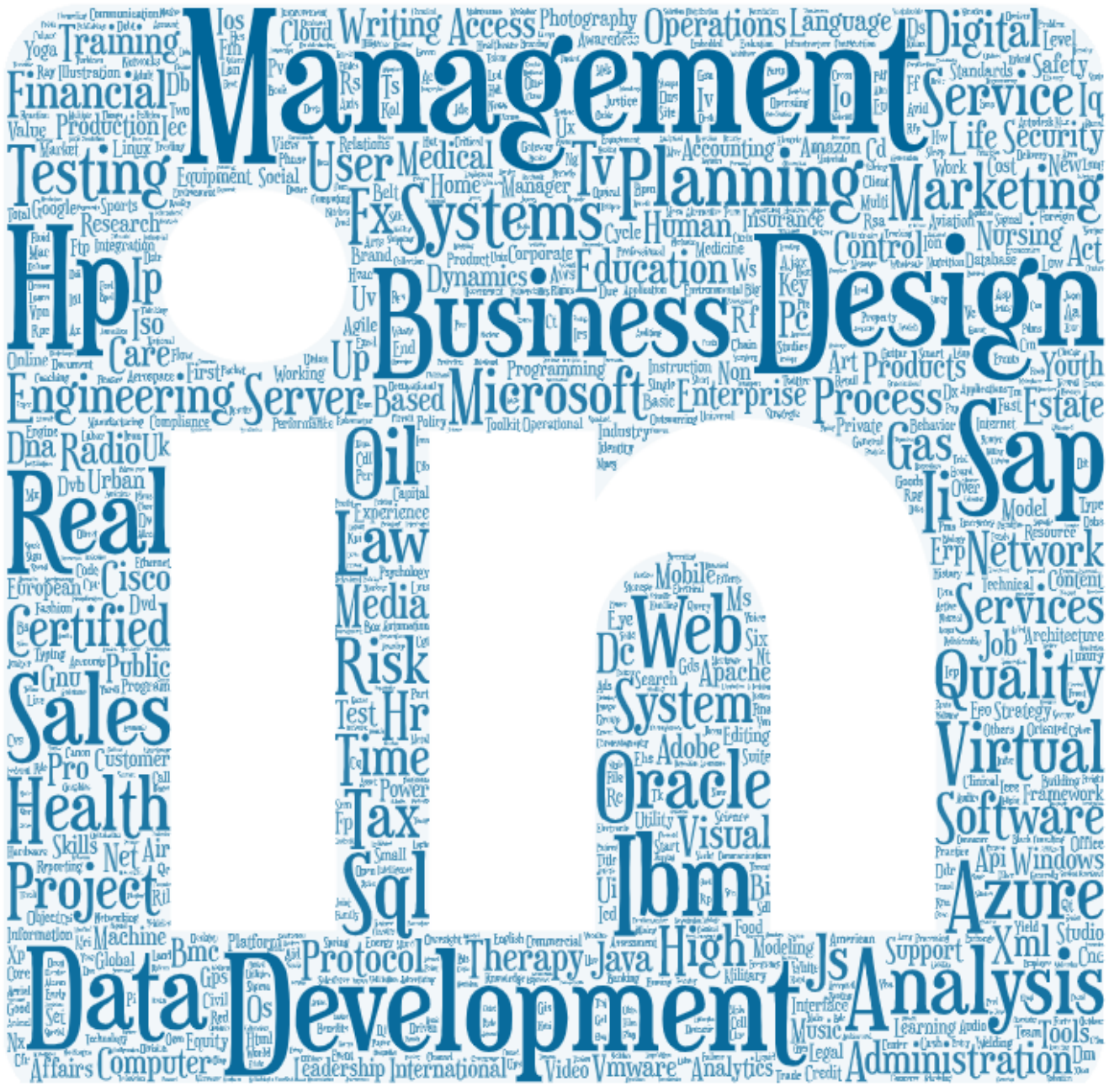

*Figure 5: crawled skills*

During the job advertisement data collection process, a total of 113,089 job postings were gathered using Python programming language and the request package. To focus on the analysis of "computer science" job ads, a filter was applied to narrow down the results which ultimately left us with 7777 job advertisements. The distribution of these job ads by year of publication is illustrated in Figure 6.

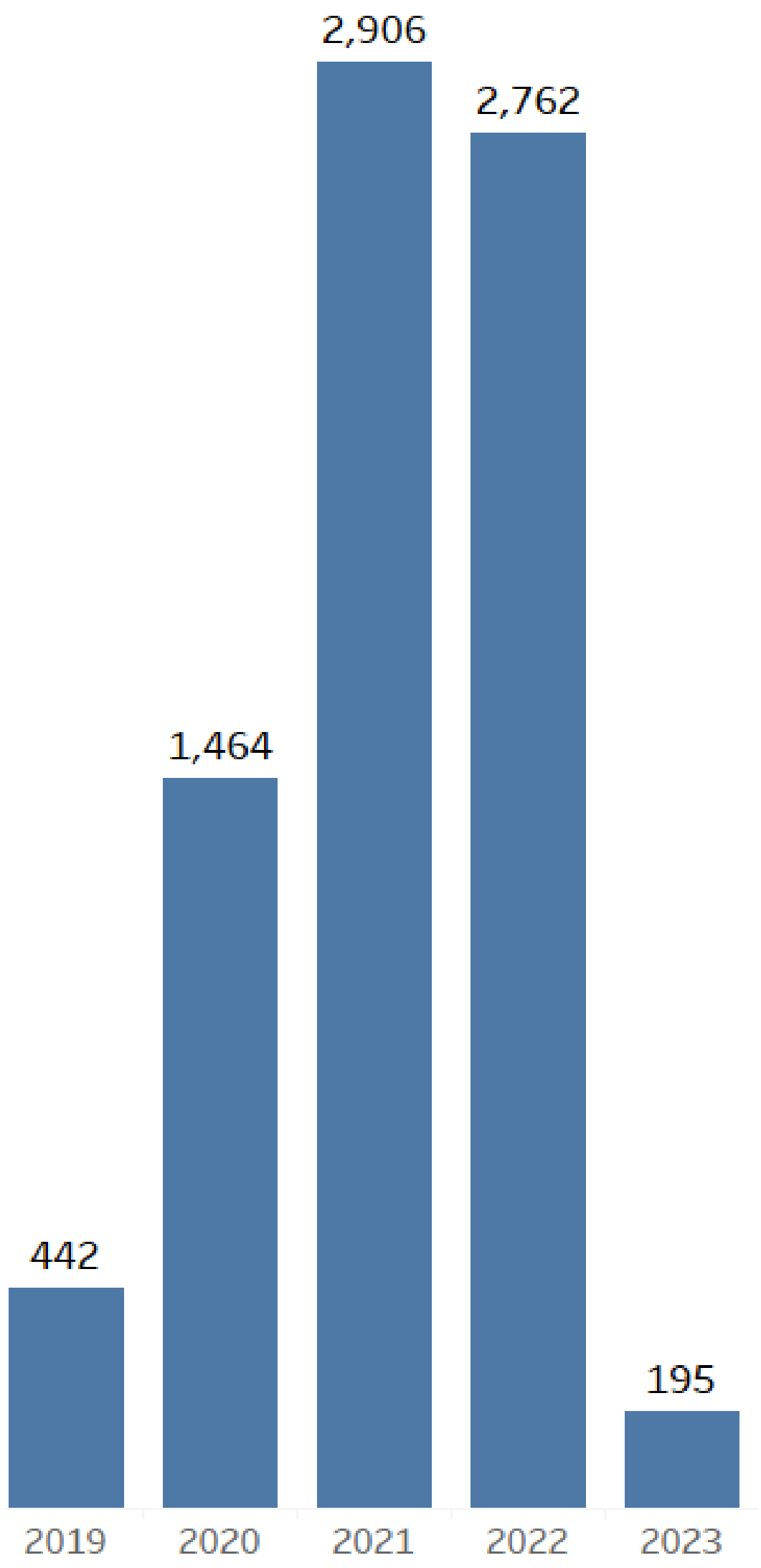

*Figure 6: Distribution of job ads per year*

## 4. 2 Data Preprocessing and Network Formation

In the process of analyzing job ads for relevant skills, a comprehensive list of 1315 unique skills emerged in the field of computer science. These skills were recorded in an ad-skill matrix. Utilizing a methodology based on common appearance in job advertisements, relationships between these skills were examined to create a skills network. As a result, 61717 distinct connections were identified, forming the foundation of the skill network illustrated in Figure 7.

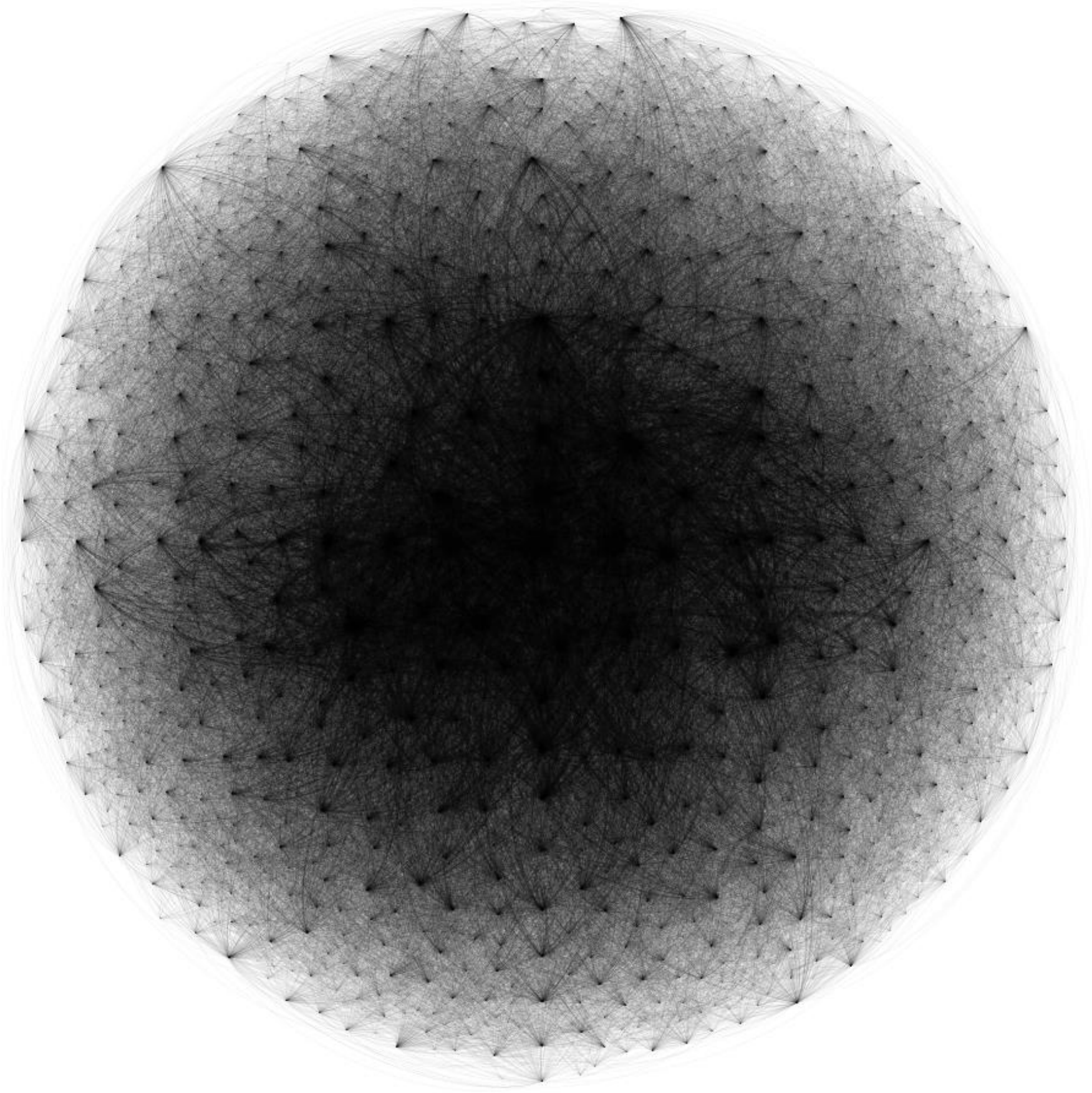

*Figure 7: skills network*

## 4. 3 Network Analysis

### 4.3.1 Analysis of Network Macro Measures

Table 3 shows the general and demographic information of the Communication Skills graph.

*Table 3: Network Macro Measures*

| Property | Value | Description |
|---|---|---|
| **Number of nodes** | 1315 | The total number of Skills available in job ads |
| **Number of Edges** | 61717 | The total number of unique relationships between Skills |
| **Average Degree** | 93.866 | The average number of unique connections between Skills |
| **Density** | 0.071 | The ratio of the number of connections created between Skills to the number of possible connections |

Based on the provided characteristics, we can make several observations about the communication network between the skills available in job ads:

- Firstly, we can see that there are a total of 1315 skills available in job ads. This indicates that there is a large pool of skills that employers are looking for in potential candidates for various jobs.
- Secondly, the total number of unique relationships between skills is 61717. This suggests that there are many different ways in which these skills are connected and related to one another. These connections likely reflect the various industries, job roles, and tasks that require a combination of skills.

- The average degree of the network is 93.866. This means that on average, each skill is connected to almost 94 other unique skills. This level of connectivity is relatively high and suggests that there is a significant amount of overlap between the skills required for different jobs.
- Finally, the density of the network is 0.071. This ratio indicates the proportion of actual connections between the skills to the total possible connections between them. A density of 0.071 suggests that the network is not completely connected but still has a significant number of connections relative to the total possible connections .

Overall, these characteristics suggest that the communication network between the skills available in job ads is complex and highly interconnected. The large pool of skills available indicates that there is a diverse range of job roles and industries being advertised. The high average degree and density of the network suggest that there is a significant overlap between the skills required for different job roles and industries.

### 4.3.2 Community Analysis

The results of four-community detection are shown in Figure 8. Communities are members of the network who have a greater desire to communicate with each other than other members of the network, which represents a specific type of Skill in the computer science job market.

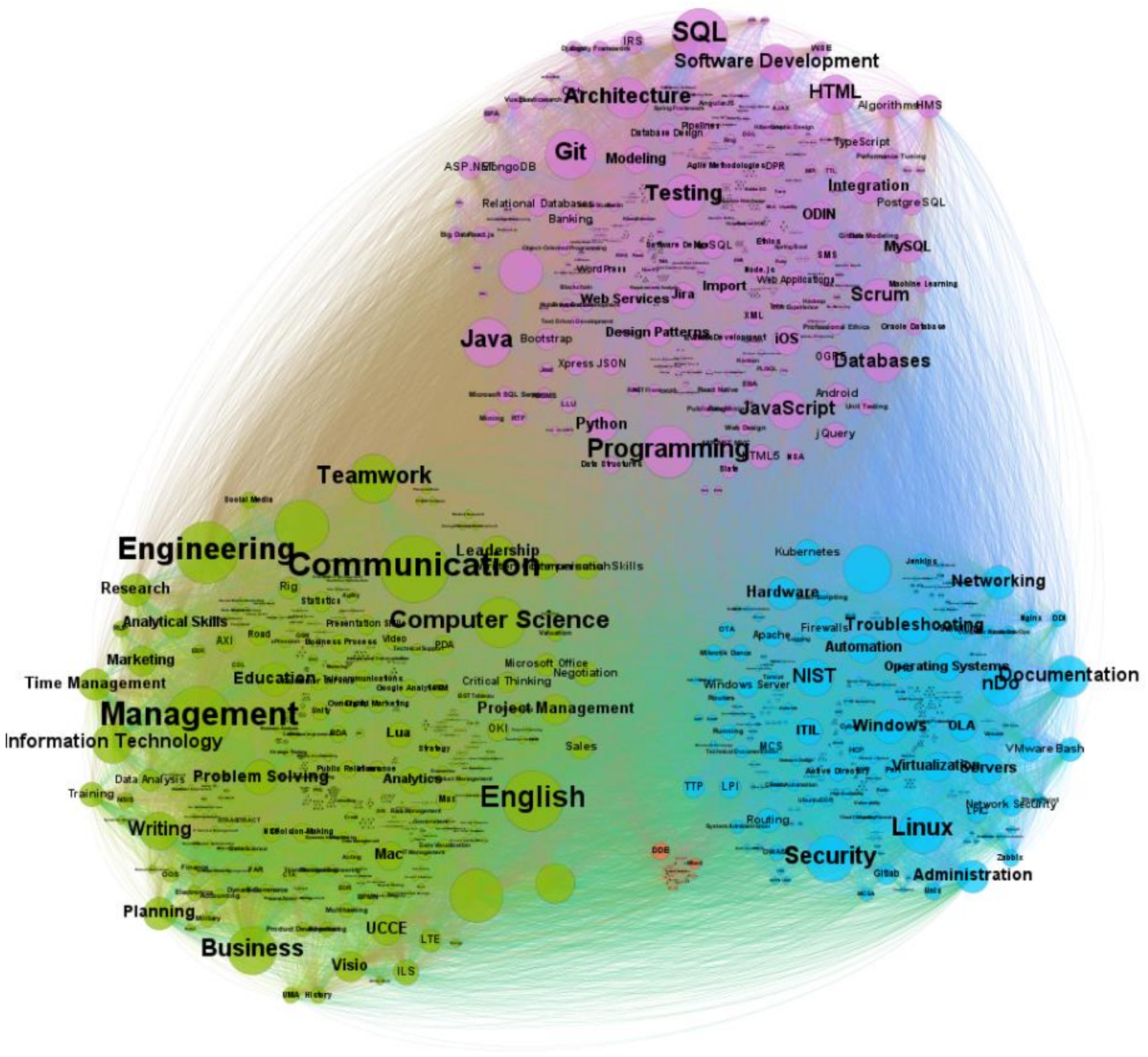

*Figure 8: Communities identified in the network*

The information obtained from detecting four communities, which includes the most important skills of each community and the proposed name, is shown in Table 4.

*Table 4: The information of four detected communities*

| Community color | Number of Members | Percent of total | The main members | Suggested name |
|---|---|---|---|---|
| • Green | ٥٢٦ | ٤٠ | Business, Communication, Education, English, Management, OLE, Planning, Problem Solving, Research, Teamwork, Time Management, and Writing | **Generalists** |
| • Purple | ٣٢٩ | ٣٣ | Administration, Automation, Documentation, Hardware, Linux, Networking, NIST, OLA, Operating Systems, Security, Servers, Troubleshooting, Virtualization, and Windows | **Infrastructure and Security** |
| • Blue | ٤٢٦ | ٢٥ | Architecture, Databases, Git, HTML, Integration, Java, JavaScript, MySQL, Programming, Python, RMAN, Scrum, Software Development, SQL, and Testing | **Software Development** |
| • Red | ٣٢ | ٢ | Altium Designer, Automotive, DDE, Embedded Software, Embedded Systems, FM, Mbed, Microcontrollers, PGA, Raspberry Pi, RS232, RTO, Simulink, SML, and SPICE | **Embedded Systems** |

Based on the skills available in the computer science job advertisements, we can divide them into four distinct communities using social network analysis. Here are the names and main members of each community:

1 .The "**Generalists**" Cluster: The members of this cluster have a wide array of Public skills such as Business, Communication, Education, English, Management, OLE, Planning, Problem Solving, Research, Teamwork, Time Management, and Writing. These areas of expertise are not restricted to any single field within computer science, and their skills can be utilized across various job roles and industries.

2 .The "**Infrastructure and Security**" Cluster: This cluster includes degrees related to infrastructure management and security, such as Administration, Automation, Documentation, Hardware, Linux, Networking, NIST, OLA, Operating Systems, Security, Servers, Troubleshooting, Virtualization, and Windows. The reason for choosing this name is that the individuals in this cluster are focused on maintaining the underlying technology infrastructure and ensuring its security, which is essential for the smooth functioning of any organization.

3 .The "**Software Development**" Cluster: This cluster includes degrees related to software development, such as Architecture, Databases, Git, HTML, Integration, Java, JavaScript, MySQL, Programming, Python, RMAN, Scrum, Software Development, SQL, and Testing. The reason for choosing this name is that the individuals in this cluster are primarily involved in designing, building, and testing software applications that serve specific business needs.

4. The "**Embedded Systems**" Cluster: This cluster includes degrees related to embedded systems, such as Altium Designer, Automotive, DDE, Embedded Software, Embedded Systems, FM, Mbed, Microcontrollers, PGA, Raspberry Pi, RS232, RTO, Simulink, SML, and SPICE. The reason for choosing this name is that the individuals in this cluster specialized in developing and implementing embedded systems, which are computer systems that are integrated into other devices or machinery.

Figure 9 shows the ratio of the number of ads that include member skills in these four communities. According to this figure, 84% of the job advertisements include the community characteristics of "general" skills, which shows the importance of these skills for employers.

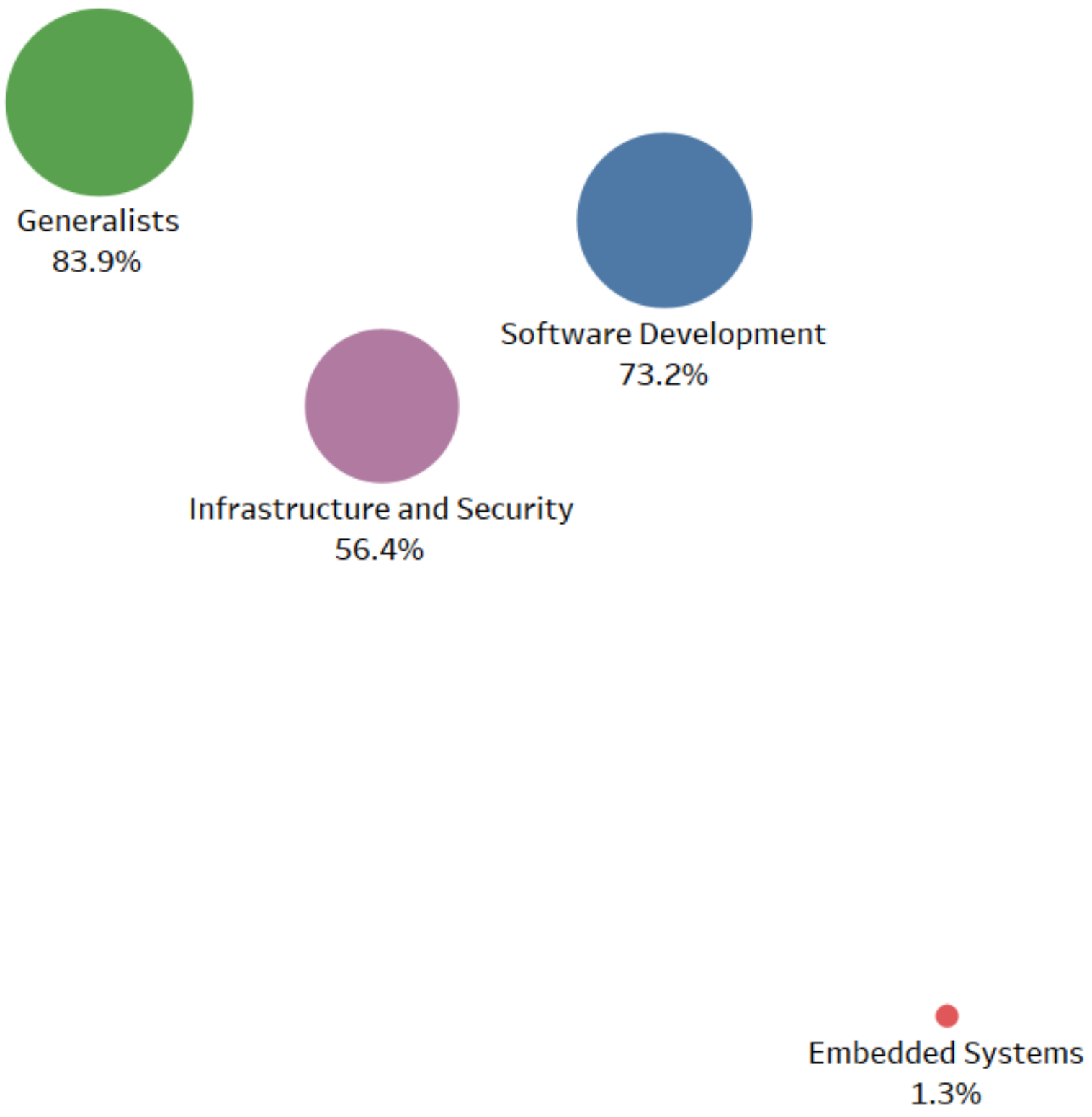

*Figure 9: the ratio of the number of ads that include member skills in these four communities*

Figure 10 provides an annual trend of the percentage of job advertisements that require skills in each of the four communities:

Generalist skills have seen a decline in demand over the past years, with approximately 87.3% of job ads in 2019 requiring these skills and dropping to 82.6% in 2023. This suggests that employers are becoming more specialized in their hiring processes and are looking for candidates who possess more specific skill sets.

Software Development skills have also seen a slight decline in demand over the past few years, from 79.4% in 2019 to 70.8% in 2023. This could be due to factors such as increasing automation and outsourced software development, which could be reducing the need for on-site or full-time software developers.

Infrastructure and Security skills, on the other hand, have seen an increase in demand over the past years, from 56.8% in 2019 to 59.1% in 2022. This could be attributed to the growing number of cyber threats and the need for companies to protect their networks and data from potential breaches.

Finally, Embedded Systems skills have remained relatively stable over the years, with only minor fluctuations in demand. This suggests that while this community remains a niche area of expertise within computer science, there is still consistent demand for skilled professionals in this field.

| Modularity Class | 2019 | 2020 | 2021 | 2022 | 2023 |
|---|---|---|---|---|---|
| Generalists | 87.3% | 87.2% | 85.2% | 81.6% | 82.6% |
| Infrastructure and Security | 56.8% | 55.1% | 55.4% | 59.1% | 56.4% |
| Software Development | 79.4% | 75.2% | 76.0% | 69.6% | 70.8% |
| Embedded Systems | 2.5% | 1.8% | 0.9% | 1.2% | 1.5% |

*Figure 10: the ratio of the number of ads that include member skills in these four communities by publishing year*

### 4.3.3 Analysis of Network Macro Measures

In a communication network of skills in job advertisements, the centrality indices can provide insights into which skills are most important for employers and job seekers .

Degree centrality measures how many connections each skill has to other skills in the network. A high degree centrality score indicates that a skill is commonly required and sought after by employers. it would be important for professionals to acquire skills with a high degree of centrality as they are likely in high demand. Betweenness centrality measures how often a skill acts as a bridge between other skills in the network. Skills with high betweenness centrality are crucial for maintaining connections and facilitating communication between different teams or departments. employers may prioritize professionals with these skills as they are key to ensuring smooth collaboration and efficient workflows.

Closeness centrality measures how quickly and easily a skill can access other skills in the network. Skills with high closeness centrality are essential for workplaces that require quick responses and decision-making. employers may seek professionals with these skills as they are capable of making informed and timely decisions when necessary. Eigenvector centrality measures the influence a skill has based on its connections to other highly influential skills in the network. Skills with high eigenvector centrality are often associated with leadership positions and can shape the direction of the company. Advertisers may value professionals with these skills as they have the potential to drive growth and success within a business.  Table 5 shows the skills with a high score in each of the centrality measures.

*Table 5: Skills with high scores in each of the centrality measures*

| Rank | Degree centrality | Betweenness centrality | Closeness centrality | Eigenvector centrality |
|---|---|---|---|---|
| 1 | Communication | Communication | Communication | Communication |
| 2 | English | English | English | English |
| 3 | SQL | SQL | SQL | SQL |
| 4 | Git | Git | Git | MICROS |
| 5 | Business | Business | Business | Git |
| 6 | Java | Teamwork | Java | Teamwork |
| 7 | Teamwork | MICROS | Teamwork | Business |
| 8 | Security | Security | MICROS | Java |
| 9 | Linux | Linux | Security | Security |
| 10 | RMAN | Java | Linux | Linux |
| 11 | Architecture | RMAN | RMAN | RMAN |
| 12 | Testing | Architecture | Architecture | Architecture |
| 13 | Documentation | Documentation | Testing | Testing |
| 14 | Databases | Testing | Documentation | Documentation |
| 15 | HTML | HTML | Databases | Databases |

According to the table, communication is ranked first across all four centrality indices, indicating its high demand and importance in the labor market. English and SQL follow closely behind, also ranking highly in all four indices. Git and business skills are also highly valued by employers, appearing in the top five for degree centrality and eigenvector centrality.

Teamwork and security skills are also considered important, ranking in the top eight for three out of four centrality indices. Proficiency in Java and Linux is also highly sought after by employers, ranking in the top ten for three out of four centrality indices.

This analysis highlights the need for both hard skills (e.g. programming languages, database management) and soft skills (e.g. communication, teamwork) in the computer science industry. Employers seem to prioritize professionals who can effectively communicate and collaborate with others while also possessing technical expertise.

Professionals seeking employment or looking to advance their careers in this field should consider acquiring these highly valued skills to improve their job prospects and marketability.

### 4.3.4 Skills Analysis

The analysis presented shows the importance of both hard skills and soft skills in the computer science industry. Soft skills such as communication, teamwork, and time management are highly valued by employers and have a high degree of centrality in the network of job advertisements analyzed. Hard skills such as programming languages, database management, and security are also important, with some experiencing increased demand over the years due to factors such as cyber threats or automation.

Employers prioritize professionals who can effectively communicate and collaborate with others while also possessing technical expertise. This highlights the need for individuals to develop a diverse set of skills to increase their marketability and job prospects.

## 5. Conclusion

the analysis of job advertisements for relevant skills in computer science reveals a complex and highly interconnected network of skills. There are 1315 unique skills available in job ads, and the total number of unique relationships between these skills is 61717. The network has a high average degree and density, indicating significant overlap between the skills required for different job roles and industries.

In this study, four distinct communities of skills were identified within the network: Generalist , Infrastructure and Security, Software Development, and Embedded Systems. The demand for Generalist skills has declined from 87.3% in 2019 to 82.6% in 2023, indicating that employers are increasingly seeking candidates with more specialized skill sets. Software Development skills have also seen a decline from 79.4% in 2019 to 70.8% in 2023 potentially due to automation and outsourced software development. In contrast, Infrastructure and Security skills have seen an increase from 56.8% in 2019 to 59.1% in 2022, while Embedded Systems skills have remained relatively stable over the years, suggesting consistent demand for skilled professionals in this field. The analysis also revealed that employers value both hard and soft skills, such as programming languages and teamwork. Communication skills were found to be the most important skill in the labor market. Additionally, certain skills were highlighted based on their centrality indices, including communication, English, SQL, Git, and business skills, among others.

Overall, the study provides valuable insights into the current state of the computer science job market and can help guide individuals and organizations in making informed decisions about skills acquisition and hiring practices. Professionals seeking employment or career advancement in the computer science industry should consider acquiring these highly valued skills to improve their job prospects and marketability. The findings also highlight the need for both hard and soft skills in the industry, including effective communication and collaboration skills. Ultimately, this study can help professionals and organizations navigate the complex and dynamic landscape of the computer science job market.